\documentstyle[psfig,times]{mn}
\begin{document}
\hsize=6truein

\renewcommand{\thefootnote}{\fnsymbol{footnote}}
\newcommand{\chandra}{{\it Chandra} }
\newcommand{\xmm}{{\it XMM-Newton} }
\newcommand{\asca}{{\it ASCA} }
\newcommand{\rosat}{{\it ROSAT} PSPC }

\title[]{Projection effects in X-ray cores of cooling flow galaxy clusters}

\author[]
{\parbox[]{6.in} {Stefano Ettori \\
\footnotesize
Institute of Astronomy, Madingley Road, Cambridge CB3 0HA \\
ESO, Karl-Schwarzschild-Str. 2, D-85748 Garching, Germany \\
settori@eso.org
}}                                            
\date{}
\maketitle

\begin{abstract}
Recent analyses of \xmm and \chandra data of the cores of X-ray 
bright clusters of galaxies show that modeling with a multi-phase
gas in which several temperatures and densities are in equilibrium
might not be appropriate. Instead, a single-phase model seems able
to reproduce properly the spectra collected in annuli from the
central region. 
The measured single-phase temperature profiles indicate a steep
positive gradient in the central 100-200 kpc and the gas density
shows a flat profile in the central few tens of kpc.
Given this observational evidence, we estimate the contribution
to the projected-on-the-sky rings from the cluster emissivity
as function of the shell volume fraction sampled. 
We show that the observed projected X-ray emission 
mimics the multi-phase status of the plasma even though 
the input distribution is single-phase.
This geometrical projection affects (i) analyses of data 
where insufficient spatial resolution is accessible,
(ii) the central bin when its dimension is comparable 
to the extension of any flatness in the central gas density profile.
\end{abstract}

\begin{keywords} 
galaxies: clustering -- X-ray: galaxies. 
\end{keywords}

\section{INTRODUCTION} 

The central cooling time of the intracluster X-ray emitting 
plasma is smaller than the Hubble time for most of the relaxed,
nearby systems requiring a flow of cooling gas in the innermost
regions to support the overlying gas and maintain the pressure
equilibrium (Fabian 1994, Allen et al. 2001). 
This phenomenon is known as cooling flow and 
appears in X-ray images as a sharply peaked central surface 
brightness and in X-ray spectra as lower temperature gas with 
high intrinsic absorption. 
    
However, a long standing debate is about the real amount of gas
deposited in the central region which can be as large as the 
mass of the central cD galaxy, $M_{\rm cool} \approx 10^{12} M_{\odot}
(\dot{M} / 100 M_{\odot} {\rm yr}^{-1}$) (Sarazin 1988, 1997).
Correlations are present between the X-ray cooling rate and the strength
of star formation in the central galaxy even if the integrated 
cooling rate up to the radius at which the cooling time becomes 
more than the Hubble time (i.e. the cooling radius, which is
about 100-200 $h_{50}^{-1}$ kpc from the centre) is orders
of magnitude larger than the rate of star formation observed only
in the inner few tens of kpc (Allen 1995; Mc Namara 1997; 
Cardiel, Gorgas \& Aragon-Salamanca 1998).
About 10 per cent of the expected mass deposited in the more massive
cooling flow clusters has been now observed in the form of molecular
gas warmed to 20--40 K by recently formed stars and located
within the inner 50 kpc radius from the cD (Edge 2001).

\begin{figure}
\psfig{figure=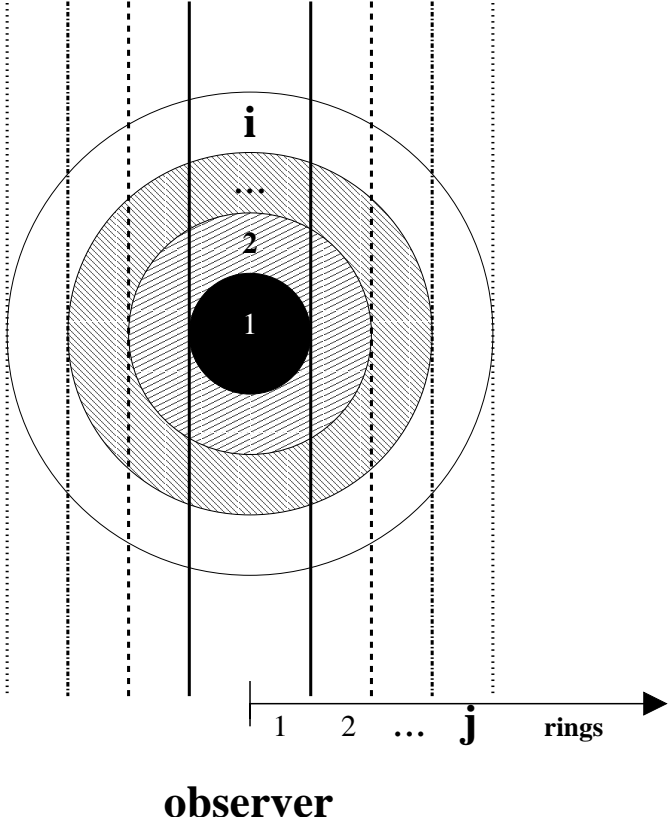,width=.5\textwidth}
\caption{Geometrical deprojection. Relation between the shells and the rings.
} \label{volfrac} \end{figure}

Recent analyses of the \xmm (Jansen et al. 2001) 
Reflection Grating Spectrometer (RGS) 
data of the cores of previously well-known
cooling flow galaxy clusters like A1795 (Tamura et al. 2001), 
A1835 (Peterson et al. 2001), Sersic 159-03 (Kaastra et al. 2001),  
do not show significant emission from gas cooling below 1--2 keV.
At the present, it is not clear what stops the gas from cooling further
or, conversely, what heats it up.
Peterson et al. (2001) and Fabian et al. (2001) discuss 
possible explanations of the observed drop in the emission from gas 
at 1--2 keV, which include heating, mixing, differential absorption
and inhomogeneous metallicity. Fabian et al. (2001) point out
how continuous or sporadic heating creates further problems,
such as, e.g., the targeting of the heat at the cooler gas and
the high total energy required (but see, e.g., Nulsen et al. 2001 and
McNamara et al. 2001).

Moreover, \chandra (Weisskopf et al. 2000)
Advanced CCD Imaging Spectrometer (ACIS) 
observation of Hydra-A (David et al. 2001)
and analyses of \xmm European Photon Imaging Camera (EPIC) 
spectra of M87, A1835 and A1795 (Molendi \& Pizzolato 2001) 
do not show a wide distribution of gas temperatures 
in the regions within the cooling radius 
(except, probably, the central 50 kpc  
where the cooling time is less than 1 Gyr)
as expected from standard multi-phase models. 
These models assume the intracluster medium
is inhomogeneous, with several comoving phases at different 
temperatures and densities, but in pressure equilibrium, at each radius.
Over a typical cooling radius of 200 kpc, the denser, cooler gas cools
out of the X-ray band before reaching the cluster centre making 
the X-ray deposition rate proportional to the radius
(see Nulsen 1986 and 1998, Thomas et al. 1987).
Furthermore, spectral analyses show strong evidence
that the gas can be modeled at the same level of accuracy (or 
sometimes better; e.g. David et al. 2001, Tamura et al. 2001)
with a single-phase instead of a multi-phase emission at each radial
annulus (again, this does not apply to the 
innermost part of the cluster where a multi-phase model appears
more appropriate).
 
Assuming that the single-phase description of the X-ray emitting plasma
is more appropriate everywhere in the cluster, 
and not considering the fate of the gas
that is cooling in this way and any other physical phenomenon
that is taking place in the cluster core,   
we have to explain why the multi-phase models
have been able to reproduce the observed spectra until now.
In this work, we focus on the role played by the geometrical projection
on the sky of the cluster core emissivity to mimic a multi-phase gas
when the underlying distribution is single-phase.
In other words, the purpose of this work is to produce a toy model
able to provide an alternative to multi-phase models and check
if the assumption of single-phase gas
can be ruled out from the results obtained
with multi-phase models.

\begin{figure*}
\hbox{
\psfig{figure=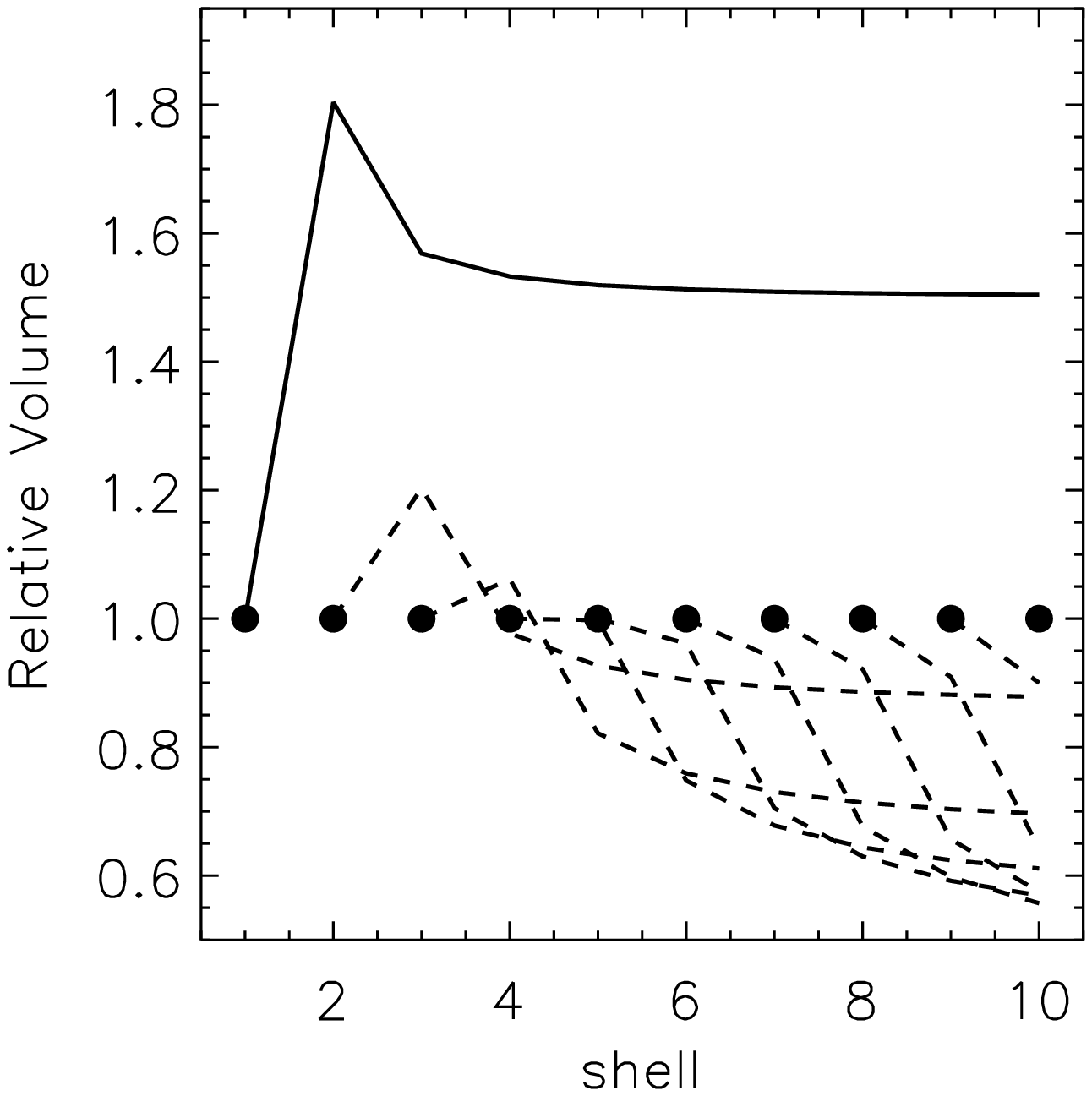,width=.5\textwidth}
\psfig{figure=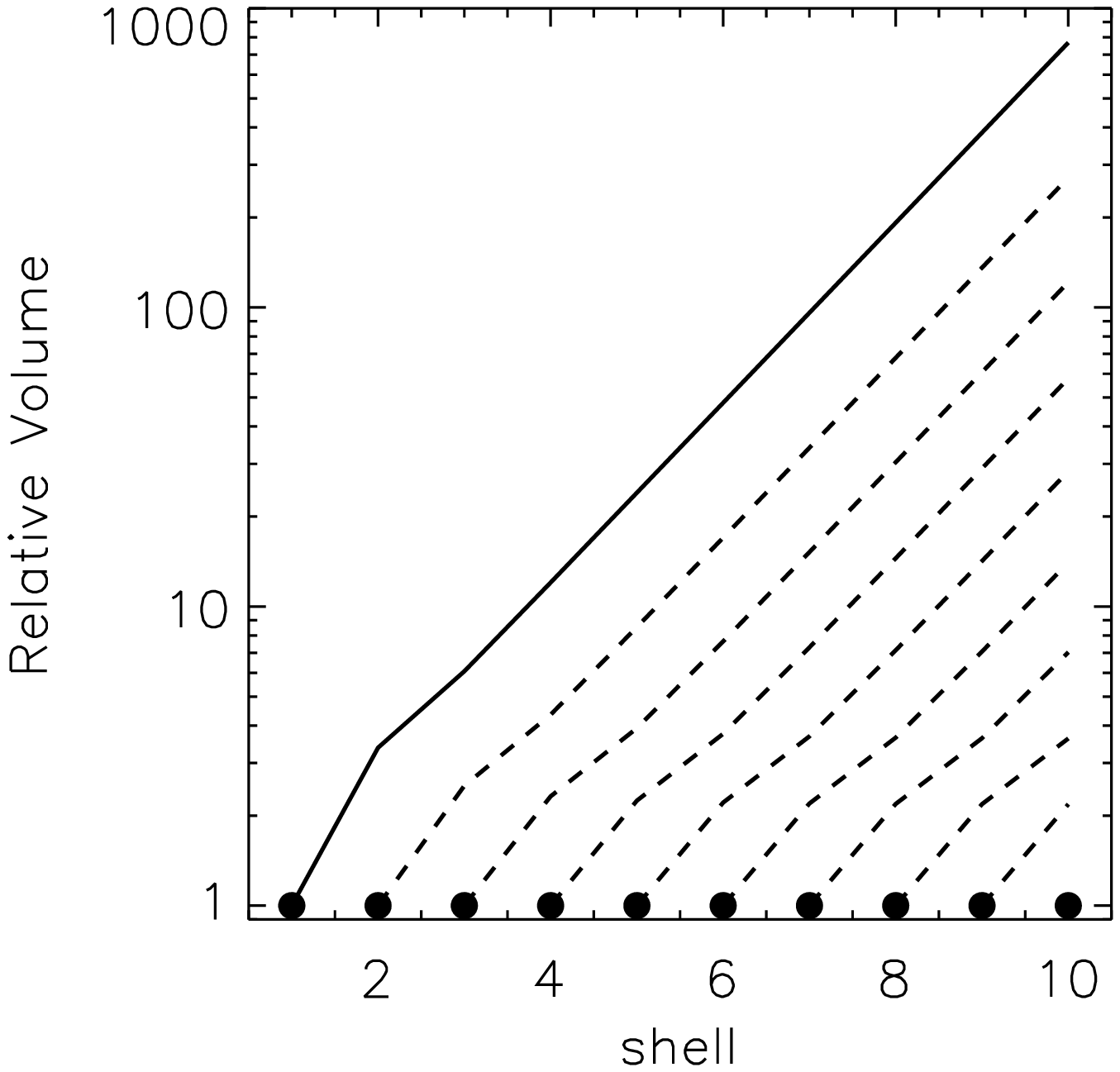,width=.5\textwidth}  }
\caption{Relative volume of the shells, normalized to the volume
of the inner shell, seen through each radial ring moving outward
from the centre. The solid line represents the inner ring.
(Left) Equally spaced bins, (right) doubling the size of the next bin.
} \label{shellvol} \end{figure*}

\section{Projection effects in the observed spectra}

We refer to Kriss, Cioffi \& Canizares (1983) and McLaughlin (1999)
for details on the calculations of the geometrical deprojection.
We define $V_{ij}$ as the amount of the volume, $V_i$, of the shell $i$
observed through the ring $j$ adopted in the spectral analysis
(cf. Fig.~\ref{volfrac}). 

In each radial ring with area $A_j$, a flux $F_j$ is measured and modeled:
\begin{equation}
F_j = \frac{\sum_{i,{\rm shell}} \epsilon_i V_{ij}}{A_j}
= \frac{\sum_{i,{\rm shell}} n_i^2 \Lambda(T_i) V_{ij}}{A_j},
\label{eq:volfrac}
\end{equation}
where $\epsilon_i$ is the emissivity in shell $i$, $n_i$ 
is the gas density and
$\Lambda(T_i)$ is the cooling function. This is proportional
to $T_i^{\alpha}$, with $\alpha$ between --0.5 and 0.5 depending on the value
of the gas temperature $T_i$ (the bremsstrahlung continuum with $\alpha 
\approx 0.5$ is predominant above 2 keV, 
whereas the emission from collisionally
excited lines mainly contribute to the total emission at lower temperatures).

Therefore, if the gas density profile in the central tens of 
kpc is flat (or not very steep),
then the integrated observed flux can be affected significantly from 
the shell volume observed.
In Fig.~\ref{shellvol}, the relative volume of the shells seen through each 
radial ring is normalized to the volume of the inner shell
and shows a peak for the inner bins 
that corresponds to the {\it second inner shell}.

Hereafter, we use equally spaced bins (or rings). It is worth noting
that this is a conservative approach, considering that any 
progressive increment in the ratio between the outer and the inner 
ring (e.g. a logarithmic scale), which is an approach generally
adopted in data analysis to increase the signal to noise
in the fainter outskirts, will imply a larger volume moving outward
(cf. Fig.~\ref{shellvol}).

\begin{figure*}
\hbox{
\psfig{figure=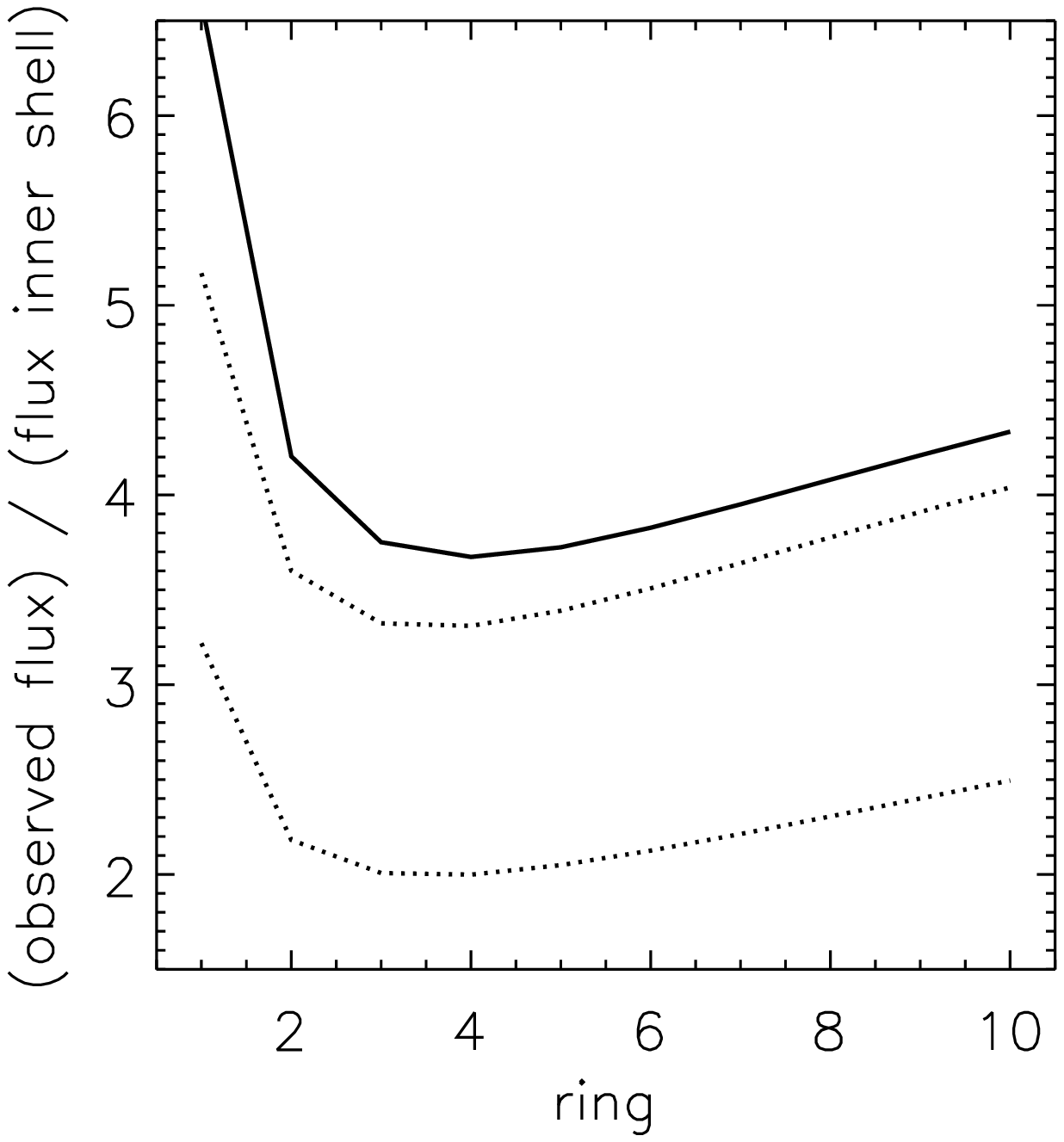,width=.5\textwidth}
\psfig{figure=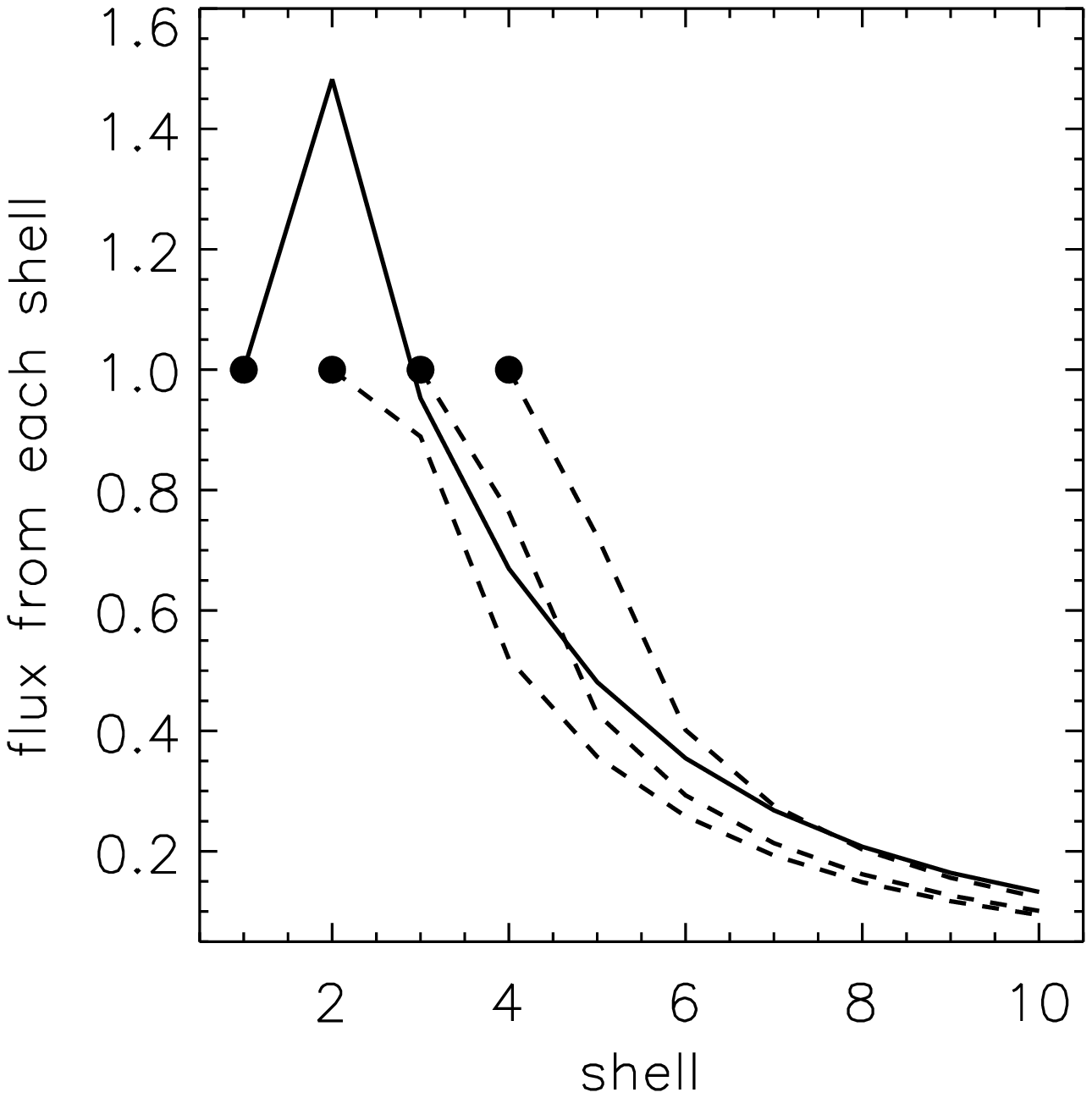,width=.5\textwidth} }
\caption{(Left) Ratio between the integrated flux observed in each ring
and the flux coming from the inner shell seen in that ring.
The solid line assumes $(\alpha, \beta, \gamma)$
=(0.5,0.4,0.2), the two dotted lines (--0.5,0.4,0.2) and (0.5,0.8,0.2)
from the top, respectively.
(Right) Differential fluxes originating from each shell and normalized
to that coming from the inner shell for the first 4
rings. The inner ring (ring 1) is represented by a solid line, the others
by dashed lines.
For a given ring, the integral of the fluxes along the shells 
(right panel) is a point
in correspondence of that ring in the left panel.
} \label{fluxdif} \end{figure*}

Given this fact, we can now estimate the differential and 
total flux measured in each shell and ring, respectively.
The null hypothesis that we investigate here is an X-ray emitting
plasma represented by a single-phase model, i.e. with a single 
temperature and density in each shell.

To model physically the intracluster medium,  
we assume a $\beta$ model (Cavaliere \& Fusco-Femiano 1976)
for the density, $n \propto (1+x^2)^{-1.5 \beta}$,
and a power law expression for the temperature:
\begin{equation}
T_i = \left\{ \begin{array}{l}
T_0 (x_i/x_0)^{\gamma} \hspace*{1cm} x_i < x_t \\
T_0 (x_t/x_0)^{\gamma} = T_t = {\rm const}  \hspace*{1cm} x_i > x_t 
\end{array}
\right.
\label{eqn:temp}
\end{equation}
It is worth noting that a single $\beta$ model is not able to represent 
over a large radial range (e.g. 2 orders of magnitude) 
the gas density profile obtained from spatial deprojection of the 
surface brightness of nearby clusters as observed with \chandra at 
high spatial resolution.
This is due to the presence in the observed profile 
of several breaks that mark significant changes in the slope
(Allen, Ettori, Fabian 2001, David et al. 2001, Ettori et al. 2001).
The $\beta$ model is however able to properly fit the gas density
in the inner region (and where no break is present) 
over 1 order of magnitude in radius
(e.g. from the \chandra observation of A1795 presented in Ettori et al. 2001, 
the gas density profile within 200 kpc is fitted  
with a $\beta-$model with $\beta = 0.38 \pm 0.01$, $r_{\rm c} = 33 \pm
2$ kpc and a $\chi^2$ of 61 with 22 degrees-of-freedom).

We have introduced two scale parameters in units of the 
core radius, $r_{\rm c}$: $x_0$ is the spatial
resolution element achievable in the spectral analysis
and gives the radial width of the central bin,
$x_t$ is the extension of the positive gradient of the gas
temperature in the core.
From \chandra and \xmm observations, we note that the spatial
resolution element is of the order of 10 kpc for nearby clusters
(at redshift 0.05, 10\arcsec correspond to 13 $h_{50}^{-1}$ kpc), 
the core radius is about 30 kpc and the radius at which the 
temperature reaches a plateau is about 200 kpc, 
generally consistent with the cooling radius.
This implies that $x_0 \approx 0.3$ and $x_t \approx 20 x_0$.
To investigate the results obtained with previous missions,
we study in the next section the case with different values
of $x_0$ and $x_t$.

Using the above models in eqn.~\ref{eq:volfrac}, and renormalizing it
with respect to the flux from the inner shell (i.e. no contribution
is considered from the outer shells), we can write
\begin{equation}
\frac{F_j}{F_{0j}} = f_j = \sum_{i=shell} v_i \ \left(
 \frac{x_i}{x_0} \right)^{\alpha \gamma} 
\ \left( \frac{1+x_0^2}{1+x_i^2} \right)^{3 \beta},
\end{equation}
where $v_i$ is the relative volume with 
respect to the volume of the inner
shell as shown in Fig.~\ref{shellvol}.

Considering equally spaced bins with dimension $x_0$, every calculation
can be simply performed in the number position of rings/shells.
We consider a region of interest extending up to 1.5 $\times x_t$. 

In Fig.~\ref{fluxdif}, we show the estimates of 
$f_j$ for $(\alpha, \beta, \gamma)$
=(0.5,0.4,0.2), typical for plasma in the core regions of nearby clusters
observed recently with \chandra, e.g. A1795 (Ettori et al. 2001) and 
Hydra-A (David et al. 2001).

From Fig.~\ref{fluxdif}, it appears that the inner ring
exhibits still significant contribution from the outer shells. In particular,
shell number 2 has a contribution by a factor of 1.48 
relative to shell number 1,
whereas numbers 3, 4, ..., 10 (see Fig.~\ref{volfrac}) have a relative
contribution of 0.95, 0.67, ..., 0.13.
The outer bins, on the other hand, are dominated by the emission 
of the central shell with contribution due to exterior shells
becoming weaker with increasing radius. 
For example, the second bin (that collects photons from all the shells 
apart from the first one and, thus, has the 2nd as innermost shell with
respect to which the relative contributions are estimated)
has a relative contribution from the 3rd shell 
of 0.89, while the 3rd, 4th, ..., 10th bins have 
a relative contribution from the next-to-the-inner shell 
of about 70 per cent. 

\begin{figure}
\psfig{figure=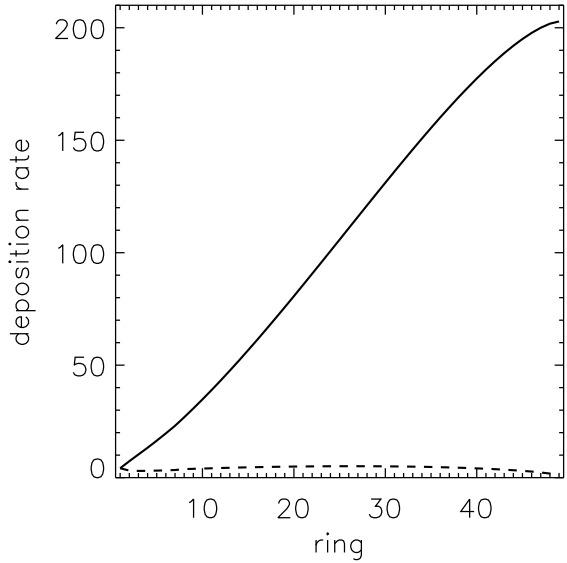,width=.5\textwidth}
\caption{Cumulative (solid line) and differential (dashed line)
distribution of the mass deposition rate, $\dot{M}$, relative
to the value in the inner shell.
} \label{mdot} \end{figure}

The combination of this differential weight in emission with the presence 
of a positive temperature gradient in the core (cf. eqn.~\ref{eqn:temp})
makes necessary the use of a multi-phase (i.e. minimun two-temperatures)
model to reproduce the observed flux.
A way to parametrize the total emission originating from a multi-phase
gas is to use the mass deposition rate, $\dot{M}$, that is defined
(from Johnstone et al. 1992) as
\begin{equation}
\frac{5}{2} \frac{T_i}{\mu m_{\rm p}} \dot{M}_{ij} = \epsilon_i V_{ij},
\nonumber \\
\dot{M}_j = \sum_i \dot{M}_{ij}.
\end{equation}
The differential and cumulative distribution of $\dot{M}$ as measured 
in each ring is plotted in Fig.~\ref{mdot} for $x_0$ = 0.3 and $x_t=$ 10.
The cumulative distribution increases with radius with a power law index
of about 1--1.1 (the dependence on $x_t$ is minimal).
This power law index is fully
consistent with both the best-fit observed values of about 1
from spatial deprojection of \rosat surface brightness 
profiles combined with \asca spectral analysis
(Allen et al. 2001) and values of 0.8 and 1.3 obtained from 
deposition rate profiles measured with spectral analysis and spatial 
deprojection, respectively, from \chandra data of A1795 
(Ettori et al. 2001).
It is worth noting that there is a slight dependence of the 
power law index upon the temperature
through the shape of the cooling function which changes from --0.5 to 0.5
moving to temperatures higher by a factor that we choose to be 
1.5 the central value (this comes from the considerations that 
(i) observations do not show temperatures lower than 1--2 keV 
in the core and (ii) bremsstrahlung is predominant at $T>$2 keV).  
This dependence makes the profile flatter in the outskirts, i.e. 
beyond the radius where we introduce the change in the shape of $\Lambda(T)$.
(e.g., when $x_0=$0.3 and $x_t=$10, the slope changes from 1.08 to 0.88).

\subsection{Changing $x_0$ and $x_t$}

In this section, we investigate how our results depend on the relative
extension of the resolution element, $x_0$, and of the temperature
gradient, $x_t$, with respect to the central flatness in the gas
density profile.

\begin{figure}
\hbox{
\psfig{figure=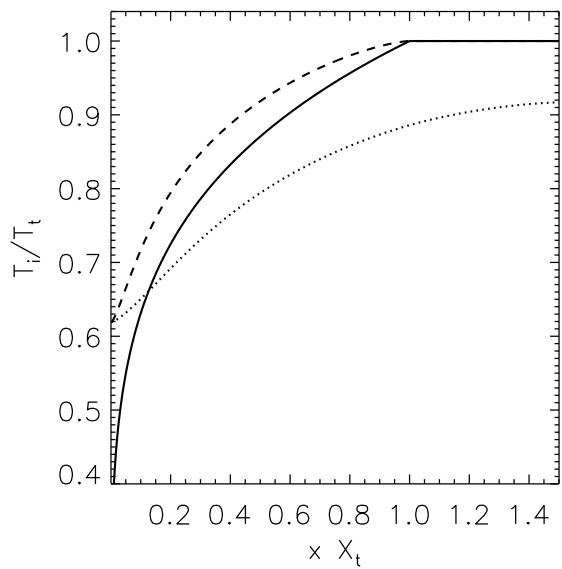,width=.5\textwidth}
} \caption{``Shell'' (solid line) and ``projected'' 
(i.e. emission-weighted in each ring; dashed line) temperature 
profiles.
The dotted line indicates the integral emission-weighted temperature
measurement, i.e. the estimated temperature value once the 
``shell'' temperature profile is weighted by the differential flux
in each shell up to a given radius. For example, at 0.5 $\times x_t$,
the ``shell'' temperature is 0.87 $\times T_t$, its projected
value at that ring is 0.92 $\times T_t$ and the estimated integrated 
value for a given detector with resolution $x_0 = 0.5 \times x_t$ 
should be 0.79 $\times T_t$.
} \label{fig:temiss} \end{figure}

Considering the weak dependence of the emissivity upon the temperature,
no significant change in the flux distribution 
is observable upon varying the values of $x_t$.
However, it is the presence of the temperature gradient that mimics
a multi-temperature distribution in the core.
The extent of this temperature gradient is proportional to $x_t/x_0$
(eqn.~\ref{eqn:temp}). Therefore, a value of $x_t$ larger than
$x_0$ is required to resolve it.
This is now routinely accessible to present-day 
missions, like \chandra and {\it Newton-XMM}.
In detail, for the well-known behaviour of the angular diameter distance
that increases up to redshifts of about 1 and then turns over,
the {\it proper} size of an object resolved   
with a 1\arcsec scale (80 per cent encircled power radius of 0.7\arcsec 
at 1 keV) by \chandra and about 20\arcsec scale 
(average of the 80 per cent encircled power radius for energies 
between 1.5 and 10.5 keV) by \xmm is
below 9 $h_{50}^{-1}$ kpc and $\sim$170 $h_{50}^{-1}$ kpc, 
respectively, at any redshift.  
In Fig.~\ref{fig:temiss}, we show the ``shell'' and projected temperature
profiles with the integral emission-weighted temperature up to a given
radius. The latter estimate gives an indication of the efficiency
in recovering the ``shell'' temperature when the spatial resolution $x_0$
is a non-negligible fraction of the extension of the temperature
gradient $x_t$.
Even if the standard cooling flow model
is not anymore a complete description of what is happening in the
cluster cores, it is worth noticing that the use of this model 
in spectral analyses for the measurement of the integral 
emission-weighted temperature within 
a fixed radius (e.g. Allen \& Fabian 1998, Ettori, Allen \& Fabian 2001)
allows to recover the ``shell'' temperature at that radius 
correcting for the presence of the cooler gas.

\begin{figure*}
\hbox{
\psfig{figure=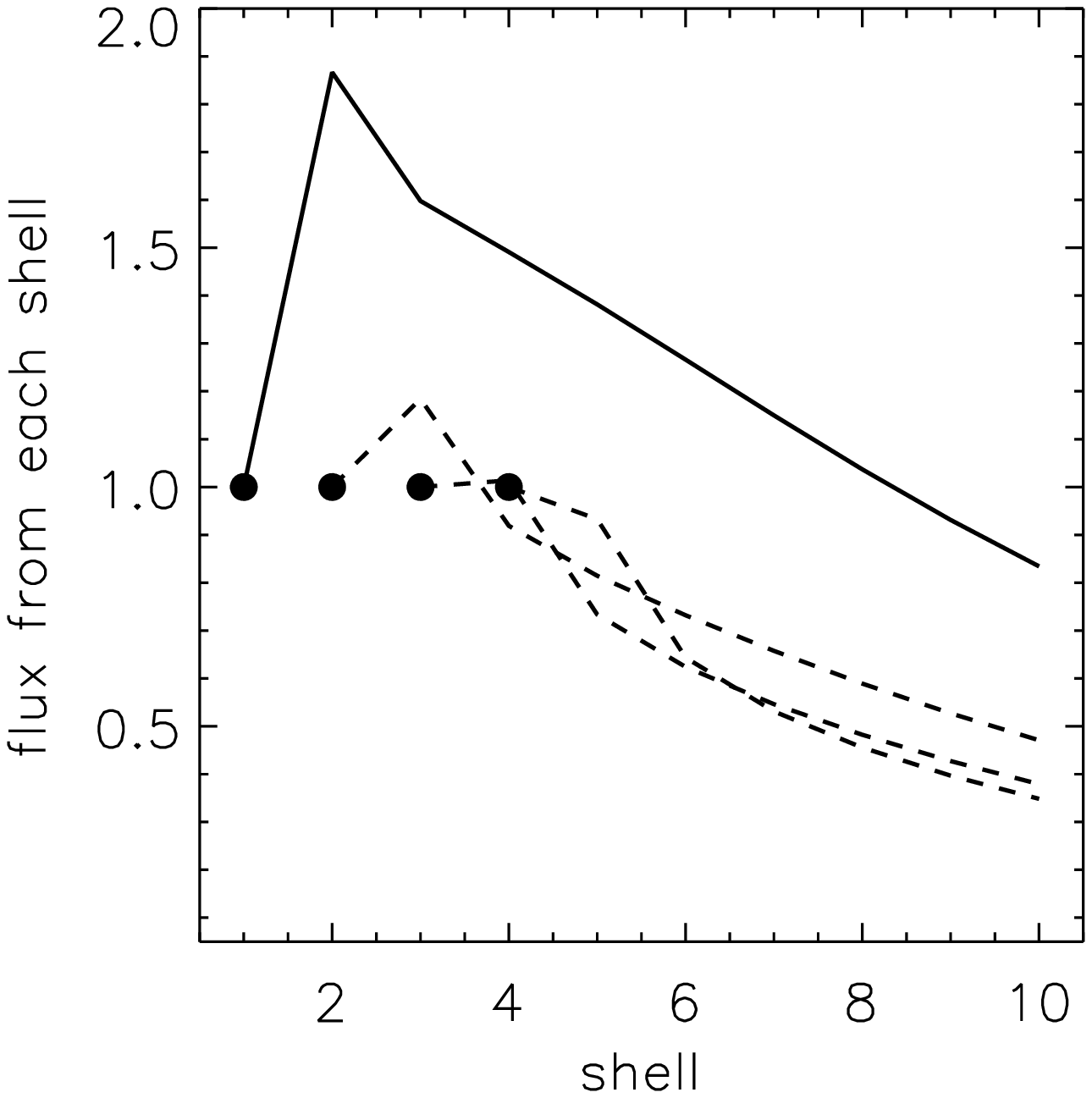,width=.5\textwidth}
\psfig{figure=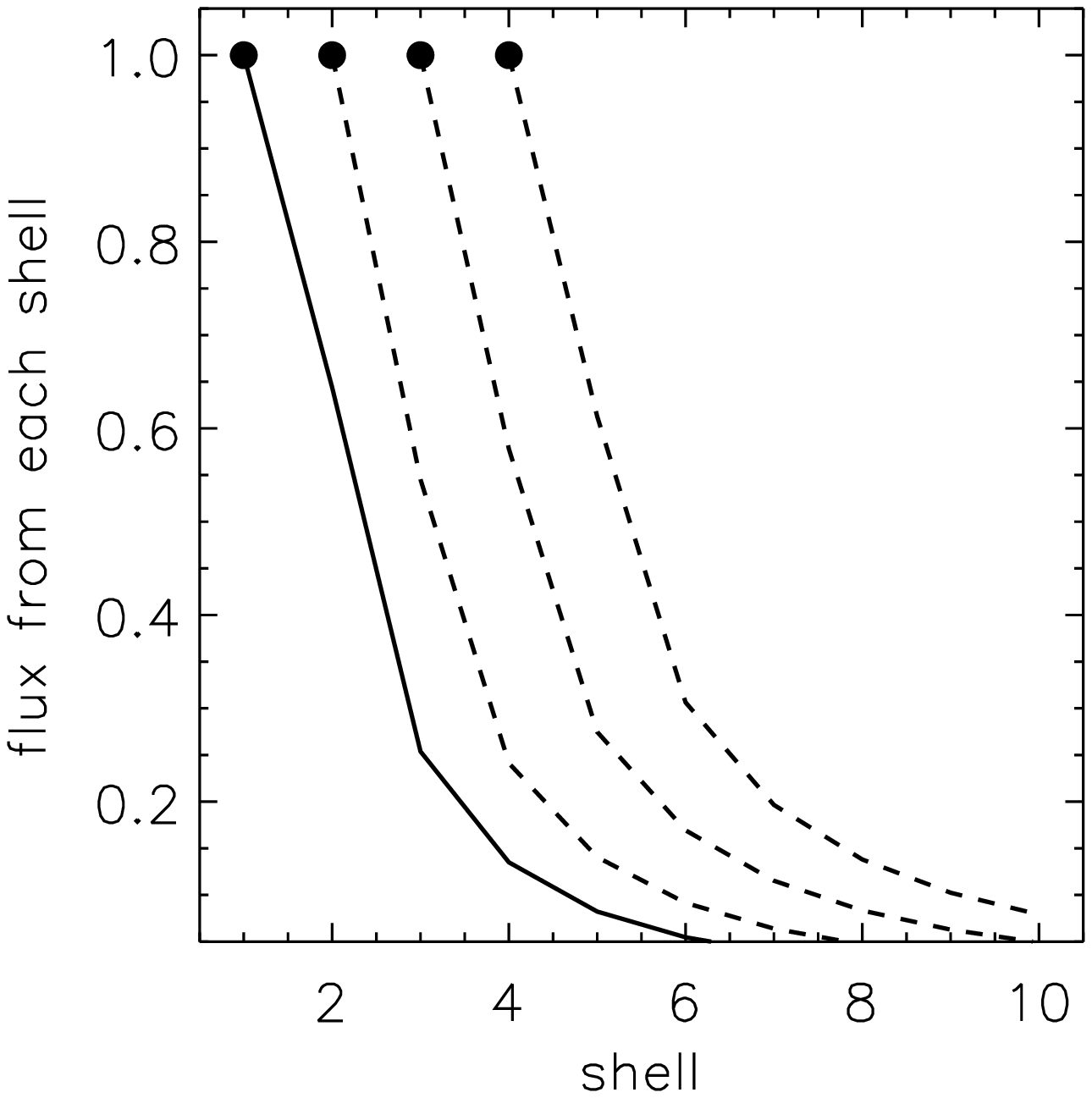,width=.5\textwidth} }
\caption{Differential fluxes in the first 4 rings
as in the right panel in Fig.~\ref{fluxdif}
but with $x_0 =$ 0.1 (left) and 1 (right) instead of 0.3.
} \label{fluxdif1} \end{figure*}

More noticeable is the influence of $x_0$. 
When $x_0$ has low values (i.e. the resolution element is smaller
than $r_{\rm c}$), more contribution from outer shells appears
as a consequence of the larger volume surveyed (see Fig.~\ref{fluxdif1}).

For example, if we enlarge $r_{\rm c}$ up to 100 kpc (or reduce
the resolution element to 3 kpc) with respect to the original
values of 30 (and 10) kpc, $x_0$ becomes 
0.1 and the contribution from shells number 2, 3, ..., 10
relative to shell number 1 in the inner bin are 1.87, 1.60, 1.49, 
..., 0.83, respectively.
The third bin still has a relative contribution from the 4th shell
(with respect to the 3rd one) of a comparable amount (1.01).
Starting with the fourth bin, the emission from inner shell (number 4)
dominates over the contribution from the next shells (e.g. 5th shell: 0.93,
6th shell: 0.64, ..., 10th shell: 0.35).
Consequently, the total observed flux relative to the flux
coming from the inner shell increases by a factor between 3.4 and 1.3 
and the deposition rate by about 2 in the first 10 rings.

On the contrary, $x_0 \ga$ 1 makes the contribution in emission from 
the inner shells dominant. When $x_0=1$, shell 2, 3, ..., 10 contributes
0.64, 0.25, ..., 0.02, respectively, to the 1st ring. 
All the outer rings have a contribution
from the second and third innermost shells of 70 and 40 per cent,
respectively. The relative observed flux is about half the one
observed for the case $x_0 =$ 0.3. 
When $x_0 > 1$ (this is the case for a given resolution element not able
to resolve any flatness in the density profile), then the contribution 
from the second (third) innermost shell is about 50--70 (20--40) per cent. 
  
Changes in $x_0$ and $x_t$ also affect the slope of the mimicked profile
of the cumulative deposition rate. 
For reasonable values of $x_t$ larger than 5, 
the larger $x_0$, the steeper the profile inside/outside 
the cooling region.
For example, the slopes vary from 0.9/0.7 when ($x_0$,$x_t$)=(0.1,5) 
to 1.1/1.0 when ($x_0$,$x_t$)=(1,10).

\section{Conclusions}

We have simulated \chandra spectra for the inner ring
with a combination of MEKAL models (Kaastra 1992, Liedahl et al. 1995)
in XSPEC (Arnaud 1996) weighted according to our 
results in Figs. ~\ref{fluxdif} and \ref{fluxdif1}.
We have then fitted both a single MEKAL model and a MEKAL 
plus intrinsically absorbed cooling flow model.
The two models provide a good fit to the data and, at the level
of the uncertainties still present, are not distinguishable
from a statistical point of view. 

We conclude that the combination of a flat gas profile in the central
tens of kpc and the volume fraction sampled of the shells 
in a projected-on-the-sky ring
explains the strong evidence of multi-phase intracluster medium whereas 
the underlying distribution is in a single-phase with a positive
temperature gradient in the core.
In particular, we show that the central bins are more affected by  
the geometrical sampling of the volume of the shells and present
a large contamination from different gas phases.
This could explain the necessity for a cooling flow
(multi-phase) component to model the emission from the
inner ring of Hydra-A (David et al. 2001)
whereas no evidence of multi-phase gas is 
spectroscopically obtained elsewhere in the cluster.
In other words, if there is evidence for a single phase gas in 
all the rings except the central one, then projection effects 
play a relevant role in contaminating the central bin contaminated  
with the overlapping shells.
These considerations support the case for a proper deprojection
of the cluster emission when the analysis of the central parts
is carried out.

\section*{ACKNOWLEDGEMENTS} The author thanks the referee, A. Edge, 
for giving suggestions in improving the presentation 
of this work and Glenn Morris for revising the original manuscript.

\end{document}